\documentclass[%
 aip,
rsi,%
 amsmath,amssymb,
 reprint,%
]{revtex4-1}
\usepackage{amsmath,bm,amssymb}
\usepackage{graphicx}% Include figure files
\usepackage{dcolumn}% Align table columns on decimal point
\usepackage{bm}% bold math
\usepackage{graphicx,epsfig}
\usepackage{amsfonts,amsmath,amssymb,amsthm,amscd}
\usepackage{color}
\usepackage{dcolumn}
\usepackage{bm}
\usepackage{array}
\usepackage{float}

\allowdisplaybreaks

\begin{document}

\preprint{AIP/123-QED}

\title{Tailored laser pulse chirp to maintain optimum radiation pressure acceleration of ions}% Force line breaks with \\

\author{F. Mackenroth}
\email{mafelix@pks.mpg.de}
\affiliation{Max Planck Institute for the Physics of Complex Systems, Dresden, Germany}%Lines break automatically or can be forced with \\
\author{S.S. Bulanov}%
\affiliation{Lawrence Berkeley National Laboratory, Berkeley, California 94720, USA}%

\date{\today}% It is always \today, today,

\begin{abstract}
Ion beams generated with ultra-intense lasers-plasma accelerators hold promises to provide compact and affordable beams of relativistic ions. One of the most efficient acceleration setups was demonstrated to be direct acceleration by the laser's radiation pressure. Due to plasma instabilities developing in the ultra-thin foils required for radiation pressure acceleration, however, it is challenging to maintain stable acceleration over long distances. Recent studies demonstrated, on the other hand, that specially tailored laser pulses can shorten the required acceleration distance suppressing the onset of plasma instabilities. Here we extend the concept of specific laser pulse shapes to the experimentally accessible parameter of a frequency chirp. We present a novel analysis of how a laser pulse chirp may be used to drive a foil target constantly maintaining optimal radiation pressure acceleration conditions for in dependence on the target's areal density and the laser's local field strength. Our results indicate that an appropriately frequency chirped laser pulse yields a significantly enhanced acceleration to higher energies and over longer distances suppressing the onset of plasma instabilities.
\end{abstract}

% \pacs{Valid PACS appear here}% PACS, the Physics and Astronomy
\keywords{laser-plasma ion acceleration, radiation pressure acceleration, high-power laser applications}%Use showkeys class option if keyword
\maketitle

\section{Introduction}
Beams of relativistic ions serve a wide range of applications from technical material science, over medical applications to even fundamental studies of high energy physics. Some of these applications particularly benefit from short, dense ion beams, not necessarily of ultra-high energy \cite{Busold_etal_2014}. Relativistic ion beams with the necessary high fluxes can be accelerated by high power lasers \cite{Mackenroth_etal_2017a}, which have undergone considerable development over the past decades \cite{Danson_etal_2015}, with several facilities breaking the Petawatt (PW) barrier already operational \cite{Hooker_etal_2006,Nakamura_etal_2017}, or in planning \cite{Major_etal_2009,Zou_etal_2015,Kawanaka_etal_2016,ELI_WhiteBook}. Consequently, the acceleration of ions to relativistic energies by high-power lasers is among the most intensely studied applications of such laser systems \cite{Mourou_etal_2006,Daido_etal_2012,Macchi_etal_2013,Bulanov_etal_2014}. 

As a result of this deep interest in laser-ion acceleration, there were several technical approaches proposed, to overcome the challenges of this application, such as the experimentally most widely studied target normal sheath acceleration (TNSA) \cite{Wilks_etal_2001,Roth_etal_2002,Mora_2003,Cowan_etal_2004,Passoni_etal_2010}, Coulomb explosion (CE) \cite{Esirkepov_etal_2002,Last_Scheck_Jortner_1997,Bulanov_etal_2002,Kovalev_Bychenkov_2003,Fourkal_Velchev_Ma_2005,Last_Jortner_2005,Murakami_Basko_2006}, hole boring (HB) \cite{Wilks_etal_1992,Naumova_etal_2009}, relativistic transparency (RT) \cite{Palaniyappan_etal_2012,Hegelich_etal_2013,Jung_etal_2013}, shock wave acceleration (SWA) \cite{Haberberger_etal_2012}, magnetic vortex acceleration (MVA) \cite{Kuznetsov_etal_2001,Bulanov_Esirkepov_2007}, standing wave schemes \cite{Mackenroth_etal_2016,Magnusson_etal_2018}, and several others to the highly efficient radiation pressure acceleration (RPA), \cite{Esirkepov_etal_2004,Bulanov_etal_2010,Henig_etal_2009,Kar_etal_2012}. In this latter regime, a thin solid density foil target is quickly ionized by the laser pulse to form a plasma, which reflects the incoming radiation and is consequently accelerated by the laser's radiation pressure. Furthermore, in the foil's rest frame the laser's frequency will appear down-shifted by a factor $2\gamma$, where $\gamma = \epsilon/m_i$ is the foil's relativistic factor, with $\epsilon$ and $m_i$ being the energy and mass of a single ion of the foil, respectively. It is thus apparent, that for relativistic foil energies ($\gamma \gg 1$) the laser light's frequency is strongly reduced leading to almost complete transfer of laser energy to the foil. In this regime, due to the relativistic time dilation, the acceleration is maintained over a long time during which the foil almost co-propagates with the laser and constantly experiences its radiation pressure. It was shown that in an ideal setting this leads to the foil reaching an energy proportional to that of the accelerating laser pulse. We note that the development of instabilities \cite{Pegoraro_Bulanov_2007} and the presence of other limiting factors, for example, laser group velocity and transverse target expansion \cite{Bulanov_etal_2015,Bulanov_etal_2016}, limit the effectiveness of the RPA. However, it was shown recently, that the laser pulse tailoring and special target engineering might compensate these limiting factors.  

The RPA was never experimentally tested in the ultra-relativistic regime due to the lack of necessary laser facilities, however  there are experimental indications that this scheme also works in the nonrelativistic regime \cite{Henig_etal_2009, Kar_etal_2012,Kim_etal_2016b,Higginson_etal_2018}. It is less stable and less effective than in the ultra-relativistic regime, mainly because the foil reflectivity is no longer perfect, but depends complicatedly on the target areal density, as well as the laser's intensity, and frequency \cite{Vshivkov_etal_1998,Macchi_etal_2009,Macchi_etal_2010,Sgattoni_etal_2012,Bulanov_etal_2012}. Including this non-trivial parameter dependence of the foil reflectivity in a one-dimensional model of its dynamics, it was demonstrated that the ion energies are optimized if for a foil of density and thickness $n_e$ and $l$, respectively, moving with a momentum $p_0(t)$ at position $\bm{x}_0(t)$, corresponding to the single-particle energy $\varepsilon(t)$, by the radiation pressure of a laser pulse with electric field envelope $E(t,\bm{x})$ and frequency $\omega_0(t,\bm{x})=2\pi/\lambda_0(t,\bm{x})$, are related via the following optimum condition \cite{Bulanov_etal_2012}
\begin{align}\label{eq:optimum}
 a_0(t,\bm{x}_0(t)) = \gamma(t) \epsilon_0(t,\bm{x}_0(t)),
\end{align}
where $a_0(t,\bm{x}) = \left|eE(t,\bm{x})\right|/(m_e\omega_0(t,\bm{x}))$ is the laser's dimensionless amplitude, the parameter $\epsilon_0(t,\bm{x})=\pi (ln_\text{e})/(\lambda_0(t,\bm{x}) n_\text{cr}(t,\bm{x}))$, introduced in \cite{Vshivkov_etal_1998}, is the target's areal density normalized to the product of the laser's wavelength $\lambda_0$ and the critical plasma density $n_\text{cr}(t,\bm{x}) = m_e \omega_0^2(t,\bm{x})/(4\pi e^2)$, where $e<0$ and $m_e$ are the electron charge and mass, respectively, and units with $c\equiv1$ are used throughout. These equations are obtained for a monochromatic laser field, whence we have to assume the pulse chirp not to be too strong, such that the frequency change can be assumed to be adiabatic. The above condition's physical meaning states that the foil should be opaque to the laser radiation at all times, in order for the required reflection to be facilitated. On the other hand, the foil density must not be too high, in order to distribute the laser energy on as few particles as possible, ensuring each individual particle experiences the largest possible energy gain. The condition (\ref{eq:optimum}) ensures the optimum compromise between these two trends. On the other hand, upon acceleration the foil becomes more and more opaque, as argued above, and hence the condition (\ref{eq:optimum}) changes over time. It was shown recently, however, how it can still be satisfied nonetheless throughout the whole acceleration process, if the laser is given an optimally tailored intensity profile \cite{Bulanov_etal_2012}.

In this paper, we study how the optimum condition (\ref{eq:optimum}) can be optimized through a tailored frequency profile, instead. We are going to demonstrate that the changes of the reflectivity can be counteracted by a complicated frequency chirp of the driving laser pulse and derive a closed analytical form of the laser's required frequency dependence. We note that the problem of the laser chirp influence on the ion acceleration was addressed in a number of papers \cite{Li_etal_2012,Liu_etal_2012,Sahai_etal_2012,Vosoughian_etal_2015}, however a systematic analytical treatment of this problem in the case of a thin foil RPA was missing. 
\section{Theory}
We begin by reformulating the optimum condition in terms of basic quantities as
\begin{align} \label{eq:optimumfield}
 \left|\mathcal{E}(t,\bm{x}_0)\right| = \gamma(t),
\end{align}
where we introduced the scaled electric field $\mathcal{E}(t,\bm{x}_0) = E(t,\bm{x}_0)/E_\text{foil}$, where $E_\text{foil} :=2\pi |e| ln_\text{e}$ is the static, one-dimensional charge separation field of the foil. Next, we note that eq.~(\ref{eq:optimumfield}) is independent of the laser's frequency and thus infer that if the laser's electric field and frequency are independent the laser's frequency cancels out of the optimum condition and, provided the field amplitude is varied appropriately, eq.~(\ref{eq:optimumfield}) is fulfilled for all frequencies. The same conclusion can be drawn from eq.~(23) of \cite{Bulanov_etal_2012}, which is independent of the laser's frequency.

\begin{figure}[t]
  \begin{center}
    \includegraphics[width=\linewidth]{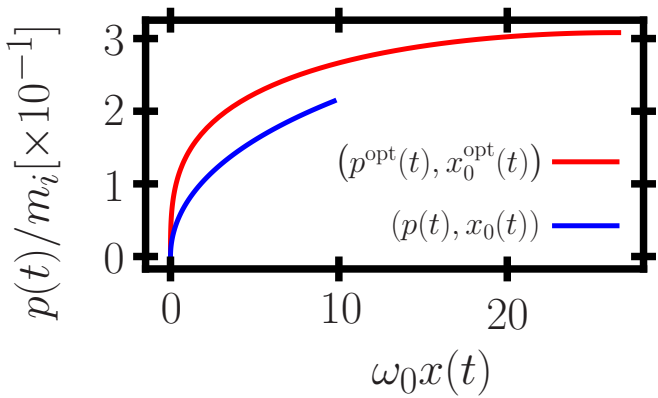}\\
    \caption{Comparison of the foil's momentum in units of an accelerated ion's mass as a function of its position in units of wavelength in the general as well as the optimized case for the parameters given in the text.}
    \label{fig:MomentumBenchmark}
  \end{center}
\end{figure}

On the other hand, the frequency still does impact the acceleration process heavily, despite the fact that the optimum condition is independent of it. To demonstrate this, we turn to the foil's equation of motion \cite{Esirkepov_etal_2004}
\begin{align}\label{eq:eom}
 \frac{d p_0}{d t} &= \frac{K\left|E(t,\bm{x}_0)\right|^2}{4\pi l n_\text{e}}\frac{\sqrt{m_i^2+p_0^2(t)}-p_0(t)}{\sqrt{m_i^2+p_0^2(t)}+p_0(t)}\\
 \frac{d x_0^\|}{d t} &= \frac{p_0(t)}{\sqrt{m_i^2+p_0^2(t)}}\nonumber \\
 K&= 2\left|\rho\right|^2 + \left|\alpha\right|^2 \nonumber ,
\end{align}
where $m_i$ is the mass of a single ion in the foil and $\rho$ and $\alpha$ are the foil's its reflection and absorption coefficient, respectively. We rewrite these equations to be expressed in the Lorentz invariant laser phase $\eta = \omega(t - x^\|(t))$, where again $x^\|(t)$ is the foil's position, to read
\begin{align}\label{eq:phaseeom}
 \frac{d p_0}{d \eta} &= \frac{d p_0}{d t} \frac{d t}{d \eta} = \frac{\left|\rho(\eta) eE(\eta)\right|^2}{eE_\text{foil}} \frac{\sqrt{m_i^2+p_0^2(\eta)}}{\sqrt{m_i^2+p_0^2(\eta)}+p_0(\eta)}\\
 \frac{d t}{d \eta} &= \frac{\sqrt{m_i^2+p_0^2(t)}}{\sqrt{m_i^2+p_0^2(t)}-p_0(t)} \nonumber .
\end{align}
Separating the variables in this equation its general solution was found to be given by \cite{Esirkepov_etal_2004}
\begin{align}\label{eq:generalmomentum}
 p_0(\eta) &= \frac12 \left(h_0 + \mathcal{D}- \frac{m_i^2}{h_0 + \mathcal{D}}\right)\\
 h_0 &:= p_0(\eta_0) + \sqrt{p_0^2(\eta_0) + m_i^2}\\
 \mathcal{D} &:= \int_{\eta_0}^\eta d\eta' \frac{\left|\rho(\eta')eE(\eta')\right|^2}{eE_\text{foil}}.
\end{align}
We continue by rewriting condition (\ref{eq:optimumfield}) as a function of the phase-dependent momentum. We find that maintaining the optimum acceleration condition is ensured by the momentum fulfilling the condition
\begin{align}\label{eq:optimummomentum}
  p^\text{opt}_0(\eta) = m_i\sqrt{\mathcal{E}^2(\eta) - 1}.
\end{align}
We immediately conclude that this condition can only be satisfied for $\left|\mathcal{E}(\eta)\right| \geq 1$, whence we have to focus on this parameter regime in the following. We do so by having our analysis only start at the time instant $\eta_0$ defined by $\left|\mathcal{E}(\eta_0)\right| = 1$, where we assumed the foil to be at position $x(\eta_0)=0$. In the following we assume the laser's intensity envelope to be given by a Gaussian of FWHM $\tau_L$, modeled by a field envelope of the form $E(\eta) = E_\text{max} \text{exp}\left[-2\log(2)(\eta/\tau_L)^2\right]$, independently of the pulse chirp. We choose to fix the pulse envelope in accordance with the above reasoning, rather than, e.g., assuming a constant value of $a_0$ or laser intensity, as we wish to model a pulsed laser field of a given intensity profile, in agreement to experimental setups. Following this model, the threshold condition can be analytically solved to give the phase at which the laser reaches the threshold of the foil transparency 
\begin{align}\label{eq:eta0}
 \eta_0 = \tau_L \sqrt{\frac{\log\left[\mathcal{E}_\text{max}\right]}{2\log(2)}},
\end{align}
such that the acceleration can only be optimized in the interval $\eta \in [-\eta_0,\eta_0]$. For all numerical examples studied below we are going to consider the acceleration only in interval $\eta \in [-\eta_0,0]$, however, as only during this interval the pulse envelope is rising allowing for an optimization of the acceleration through a pulse chirp, as argued below. Furthermore, here we consider a long laser pulse with $\tau_L = 10^2/\omega_0$. In order to test the improved ion acceleration regime, we numerically integrate eq.~(\ref{eq:optimummomentum}) and compare it to a full numerical solution of the system (\ref{eq:phaseeom}). Furthermore, as we additionally wish to highlight the enhancement in the accelerated foil's position $x_0(\eta)$, we additionally solve its equation of motion
\begin{align}\label{eq:phaseposition}
 \frac{d x_0}{d \eta} &= \frac{d x_0}{d t} \frac{d t}{d \eta} = \frac{p_0(t)}{\sqrt{m_i^2+p_0^2(t)}-p_0(t)},
\end{align}
where we made use of the relativistic velocity relation $dx_0/dt = p_0(t)/\sqrt{m_i^2+p_0^2(t)}$. In order to analyze the solutions of the respective equations of motion for the foil's momentum and position in the general as well as the optimized case we study a foil of density $n_e = 10^{24}\, \text{cm}^{-3} \approx 574 n_\text{cr}$ for radiation of $800$ nm wavelength, and thickness $l=10$ nm accelerated by a moderately relativistic laser with intensity $I = 1.5\times 10^{21} \text{W}/\text{cm}^2$. Furthermore, as we wish to analyze the efficiency enhancement as a function of space-time, we plot the foil's momentum in the laboratory frame as a function of its position by means of a parametric plot with the coordinates $\left(p_0(\eta)/m_i,\omega_0 x_0(\eta)\right)$ (s.~fig.~\ref{fig:MomentumBenchmark}). We find a significant enhancement of the foil's momentum, when accelerated according to the optimum condition $p_0^\text{opt}$, derived from eq.~(\ref{eq:optimummomentum}).
We furthermore note that while we only consider the acceleration up to the phase value $\eta=0$, the physical time extends to values larger than $0$, for the simple reason that the non-trivial foil displacement $x_0(t)$ leads to a non-trivial relation between phase $\eta$ and time $t=\eta + x_0(t)$. In this respect, we also note that since $p_0^\text{opt}(t) > p_0(t)$ it will be $x_0^\text{opt}(t) > x_0(t)$ as reflected in the fact that $\eta=0$ is reached at larger displacements $x_0^\text{opt}(\eta)$ in the optimized regime. We note that this can be interpreted as the optimized case leading to a reduction of the acceleration length, i.e., the same ion energy can be reached over smaller distance. This effect is clearly inferable from the fact that in the optimized case the ion momentum at any given foil position is is higher than in the unoptimized case, indicating that the optimized pulse chirp needs significantly less distance to accelerate ions to a certain energy. Usually the acceleration length is limited by either the Rayleigh length or the transverse expansion of the target, or both. In the case of transversely flat-top (e.g., super-Gaussian) laser pulses, which are often suggested to be employed to produce quasi-mono energetic ion beams, the acceleration length is limited by the fact that such pulses do not propagate without changing their transverse shape. Thus, any technique that allows for reaching some ion energy over shorter distance is bound to optimize the acceleration process.

\begin{figure}[t]
  \begin{center}
    \includegraphics[width=\linewidth]{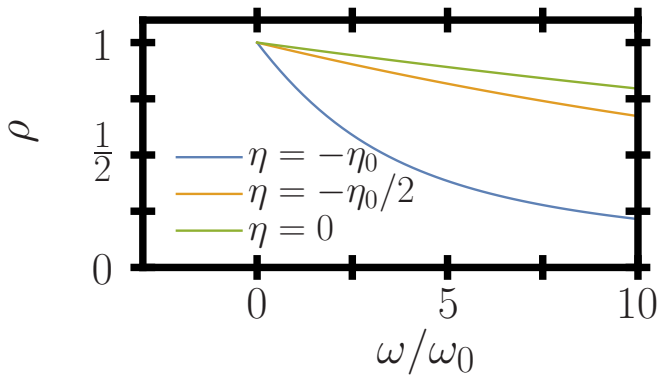}
    \caption{Reflection coefficient as a function of the laser's frequency at various phases of the driving laser pulse for a foil of $10$ nm thickness and a density of $n_e = 10^{23} /\text{cm}^3$ ($\approx 50 n_\text{cr}(\omega_0=1.55 \text{eV})$).}
    \label{fig:reflection}
  \end{center}
\end{figure}

Having established the improved performance of eq.~(\ref{eq:optimummomentum}) we continue to discuss how a frequency chirp can be used to maintain the optimum condition eq.~(\ref{eq:optimum}). To this end, we require the reflection coefficient entering eq.~(\ref{eq:phaseeom}). A foil's reflection coefficient depends on the laser's and foil's parameters in its rest frame ($\gamma \approx 1$) according to \cite{Vshivkov_etal_1998}
\begin{align}\label{eq:generalreflection}
 \rho(\eta) = \frac{\epsilon_0(\eta)}{a_0(\eta)} \left(\frac{\left(\left[\Delta^2 - 1\right]^2 + 4a_0^2\right)^{\frac12} + \Delta^2 - 1}{\left(\left[\Delta^2 - 1\right]^2 + 4a_0^2\right)^{\frac12} + \Delta^2 + 1}\right)^{\frac12},
\end{align}
where we defined the difference $\Delta^2(\eta) := a_0^2 (\eta)- \epsilon_0^2(\eta)$. Transforming this relation back into the laboratory frame amounts to the replacement $\epsilon_0(\eta) \to \gamma(\eta) \epsilon_0(\eta)$. Hence, from eq.~(\ref{eq:optimum}) we read off that both in the foil's rest frame and the laboratory frame the optimum condition can be used to simplify the reflection coefficient. Consequently, in this work we can always use
\begin{align}\label{eq:simplifiedreflection}
 \rho(\eta) = \left(\frac{\left(1 + 4a_0^2\right)^{\frac12} - 1}{\left(1+ 4a_0^2\right)^{\frac12} + 1}\right)^{\frac12}.
\end{align}
Assuming then properties of a typical foil of $10$ nm thickness and a density of $n_e = 10^{23} /\text{cm}^3$, approximately corresponding to $50$ times the critical density for an optical laser beam of $800$ nm wavelength, driven by a laser with the above assumed Gaussian temporal profile and peak intensity $I=10^{21}\, \text{W}/\text{cm}^2$ we find a decisive dependency of the reflection coefficient on the incident frequency (s. fig.~\ref{fig:reflection}). We can also read off that this dependency is more pronounced the in the laser's rising edge where $a_0$ is lower, indicating that the laser frequency gives the finest tunability during the starting phase of RPA.
Based on these results, it is apparent that even though eq.~(\ref{eq:optimummomentum}) is formally independent of the laser's frequency, an appropriate pulse chirp can still be used to tailor the reflection coefficient and hence the overall acceleration. This approach is complementary to the tuning through an optimized intensity profile \cite{Bulanov_etal_2012}, which is aimed at the ultra-relativistic regime. In contrast, the here presented frequency optimization is most apt to steer and stabilize the commonly highly unstable initial phase of radiation pressure acceleration. To find an optimal pulse chirp, a differential approach is favorable: Provided we can match the foil's momentum to its ideal value at a given time instant, from that time on the optimum condition can be enforced on its differential equation of motion (\ref{eq:optimummomentum}). The optimum momentum, fulfilling eq.~(\ref{eq:optimumfield}) perpetually, on the other hand, changes as
\begin{align}\label{eq:optimummomentumchange}
 \frac{d p^\text{opt}}{d\eta} = m_i \frac{\mathcal{E}'(\eta)\mathcal{E}(\eta)}{\sqrt{\mathcal{E}^2(\eta) - 1}}.
\end{align}
This differential change can be matched to eq.~(\ref{eq:phaseeom}) provided the reflection coefficient is given by
\begin{align}\label{eq:optimumreflection}
 \rho(\eta) = \sqrt{\left(\frac{1}{\sqrt{\mathcal{E}^2(\eta) - 1}}+\frac{1}{\mathcal{E}(\eta)}\right)\frac{m_i\mathcal{E}'(\eta)}{\left|eE(\eta)\right|}},
\end{align}
where we additionally assumed eq.~(\ref{eq:optimummomentum}) to be fulfilled and the foil to be initially at rest $p(\eta_0)=0$. Equating eqs.~(\ref{eq:simplifiedreflection},\ref{eq:optimumreflection}) and solving for the laser frequency numerically will be a benchmark for our analytical results. We note that provided the optimal reflection coefficient (\ref{eq:optimumreflection}) is maintained throughout the acceleration process, the foil ions' momentum will develop according to eq.~(\ref{eq:optimummomentumchange}) and be given by eq.~(\ref{eq:optimummomentum}). Additionally, we see that eq.~(\ref{eq:optimumreflection}) is only meaningfully defined in the regime $\mathcal{E}'(\eta) \geq 0$, i.e., on the pulse's rising edge. This is physically due to the fact that according to eq.~(\ref{eq:optimumfield}) on the field envelope's decreasing edge a decreasing $\gamma(\eta)$ would be required to maintain the optimum condition, which would correspond to deceleration instead of the desired acceleration. We thus have to limit our analysis to the first half of the pulse, discriminated by $\mathcal{E}'(\eta) \geq 0$. We note that the results of Ref. \cite{Bulanov_etal_2012} indicate that the optimal pulse profile has only the rising edge.

In accordance with the technical development we are going to consider high-intensity laser pulses defined by the condition $a_0\gg1$ and remind ourselves that the foil will be driven in a parameter regime close to the optimum condition (\ref{eq:optimum}), indicating that $a_0(\eta) \approx \gamma(\eta) \epsilon_0(\eta)$, where we did not yet assume eq.~(\ref{eq:optimum}) to hold exactly. We can then study the acceleration process for the case of an ultra-relativistic foil $p_i \gg m_i$, as is common in investigations of the RPA scheme, and in addition in the foil's not commonly considered beginning, still non-relativistic acceleration phase $p_i \sim m_i$. 
\subsection{Non-relativistic foil motion}
We begin studying the motion of a non-relativistic foil $1-\gamma(\eta) \ll 1$, which, according to eq.~(\ref{eq:optimumfield}) translates to the condition $\left|\mathcal{E}(\eta)\right|\sim1$. From eqs.~(\ref{eq:optimum}) we deduce that in this case close to the optimum drive regime the foil's and laser's parameters are linked by $\epsilon_0 \approx a_0$. Furthermore, from eq.~(\ref{eq:eta0}) we see that in the regime $\mathcal{E}\sim 1$ the acceleration will be confined to phases $\eta_0 \ll \tau_L$, such that the field will not be strongly changing $\mathcal{E}'(\eta) \ll \mathcal{E}(0)/\tau_L$. From eq.~(\ref{eq:optimumreflection}) we thus see that the optimal reflection coefficient has to be rather small, which is achievable for large laser frequencies (compare fig.~\ref{fig:reflection}). This in turn, however, implies small values of $a_0$, even for high laser powers. In order to corroborate this conjecture, we need to find an analytic expression for the laser's frequency structure ensuring that the foil's reflection coefficient from eq.~(\ref{eq:simplifiedreflection}) is matched to its optimal reflection derived in eq.~(\ref{eq:optimumreflection}). To find such a solution for the optimized laser frequency, we again solve the foil's equation of motion (\ref{eq:phaseeom}) through separation of variables and find that, when neglecting absorption in the foil, in the non-relativistic regime the momentum of a foil initially at rest is given by
\begin{align}
 p_0(\eta) &= \int_{\eta_0}^\eta d\eta' \frac{\left|\rho(\eta')eE(\eta')\right|^2}{eE_\text{foil}}
\end{align}
The same result can be found expanding solution (\ref{eq:generalmomentum}) to lowest order in $p_0/m_i$. From eq.~(\ref{eq:optimummomentum}) we deduce that the phase dependent reflection coefficient required to meet the optimum condition is given by
\begin{align}\label{eq:NR_optimumreflection}
 \rho(\eta) = \sqrt{\frac{m_i \mathcal{E}'(\eta)}{\left|eE(\eta)\right|\sqrt{\mathcal{E}^2(\eta) - 1}}},
\end{align}
which is equivalent to approximating the solution of eq.~(\ref{eq:optimumreflection}) in the regime $\sqrt{\mathcal{E}^2(\eta) - 1} \ll \mathcal{E}(\eta)$. Equating this to the approximation (\ref{eq:simplifiedreflection}) we find that an appropriate laser frequency chirp can ensure condition (\ref{eq:optimum}) to be satisfied in the regime $\mathcal{E}\sim1$ provided it holds
\begin{align}\label{eq:NRoptimum}
 \omega_\text{NR}^\text{opt}(\eta) = \frac{eE(\eta)}{m_e} \left(\frac{\left|eE(\eta)\right|\sqrt{\mathcal{E}^2(\eta) -1}- m_i \mathcal{E}'(\eta)}{\sqrt{m_i \mathcal{E}'(\eta)\left|eE(\eta)\right|\sqrt{\mathcal{E}^2(\eta) -1}}}\right).
% \omega(\eta)= \frac{2\left|eE(\eta)\right|}{m_e }\left(1- \sqrt{\frac{m_i \mathcal{E}'(\eta)\mathcal{E}(\eta)}{2\pi e^2 ln_e\sqrt{\mathcal{E}^2(\eta) - 1}}}\right).
\end{align}
\begin{figure}[t]
  \begin{center}
    \includegraphics[width=\linewidth]{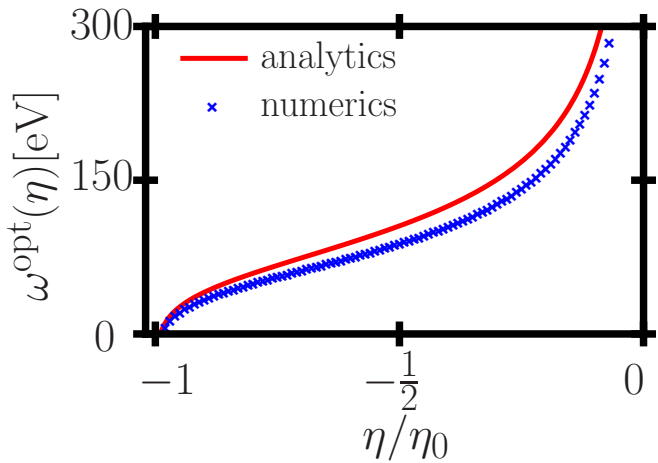}
    \caption{Optimal laser frequency in the non-relativistic model case (parameters in the text) according to the analytical solution eq.~(\ref{eq:NRoptimum}) (solid red) in comparison to the numerical solution of eqs.~(\ref{eq:simplifiedreflection},\ref{eq:optimumreflection}) (blue crosses).}
    \label{fig:NRoptimum}
  \end{center}
\end{figure}
Comparing this analytical expression to the numerically consistent solution of eqs.~(\ref{eq:simplifiedreflection},\ref{eq:optimumreflection}) for a foil of $10\, \mu$m thickness and $n_e = 9\times10^{23}\, \text{cm}^3 \approx 520\, n_\text{crit} $ accelerated by a laser pulse of intensity $I=10^{21}\, \text{W}/\text{cm}^2$ and $\tau_L = 10^3/\omega_0$ we find good agreement with the outlined derivation (s. fig.~\ref{fig:NRoptimum}). We also find our conjecture confirmed that the required laser frequencies are several hundreds of eV, as issued above.
\subsection{Ultra-relativistic foil motion}
We now turn to studying the motion of an ultra-relativistic foil $\gamma(\eta) \gg 1$, which, according to eq.~(\ref{eq:optimumfield}) translates to the condition $\left|\mathcal{E}(\eta)\right|\gg1$. In this regime, as can be deduced from eq.~(\ref{eq:eta0}) we see that the acceleration will be occurring over the whole phase interval $\eta \in \left[-\tau_L,\tau_L\right]$. Consequently, in contrast to the previously studied non-relativistic case, the field's derivative can be estimated to be of the order $\mathcal{E}'(\eta) \sim \mathcal{E}(\eta)/\tau_L$. Hence, we can estimate from eq.~(\ref{eq:optimumreflection}) that in this regime the optimal reflection coefficient will be of the order $\rho \lesssim \sqrt{2m_i/\left|eE(\eta)\right|\tau_L}$. This indicates that for ultra-relativistic foil motion in the pulse center, i.e., for largest field strengths, the optimally matched reflection coefficient again has to be small, which in the present scheme is achievable by large frequencies. On the other hand, from eq.~(\ref{eq:simplifiedreflection}) one infers that the maximal reflection coefficient $\rho\to1$ is reached for $a_0 \to \infty$, corresponding to the low-frequency limit $\omega \to 0$. As a result, in the beginning phase of the acceleration the reflection coefficient can still be small, facilitating the use of optical laser frequencies. From this result, however, we infer a further restriction: The optimum condition (\ref{eq:optimumfield}) can only be maintained up to the phase instant, where the value of the required optimally matched reflection coefficient from eq.~(\ref{eq:optimumreflection}) exceeds the maximum achievable value $\rho_\text{max}$. This phase instant is implicitly defined by the condition
\begin{align}\label{eq:derivativecondition}
 \mathcal{E}'(\eta) \leq \frac{\left|e E_\text{foil}\right|}{2m_i}.
\end{align}
Assuming again the Gaussian field shape $E(\eta) = E_\text{max} \text{exp}\left[-2\log(2)(\eta/\tau_L)^2\right]$, in the pulse's rising edge the scaled field's derivative is maximal at the phase instant $\tilde{\eta}= - \tau_L/2\sqrt{\log(2)}$, whence for the specified pulse shape we can rewrite eq.~(\ref{eq:derivativecondition}) as a maximal condition for the pulse duration in the form
\begin{align}\label{eq:durationcondition}
 \tau_L\geq \frac{4m_i\sqrt{\log(2)}}{\left|e E_\text{foil}\right|}\mathcal{E}_\text{max} \text{e}^{-\frac{1}{2} },
\end{align}
which ensures that the optimally matched reflection coefficient can always be matched by eq.~(\ref{eq:simplifiedreflection}). Since we fix the intensity profile, the change in laser frequency leads to $a_0$ changing its value accordingly by a couple of orders of magnitude. Consequently, for ultra-relativistic foil motion, the matching between eq.~(\ref{eq:simplifiedreflection}) and the foil's optimal reflection coefficient eq.~(\ref{eq:optimumreflection}) requires to again solve eq.~(\ref{eq:phaseeom}) by separating the variables and integration. The result is formally the same expression for the foil's momentum as in the non-relativistic case but, from eq.~(\ref{eq:optimummomentum}) we deduce that in the case $\gamma\gg1$ the phase dependent reflection coefficient required to meet condition (\ref{eq:optimummomentumchange}) is given by
\begin{align}\label{eq:UR_optimumreflection}
 \rho(\eta) = \sqrt{\frac{2 m_i E'(\eta)}{eE^2(\eta)}}
\end{align}
which is equivalent to approximating eq.~(\ref{eq:optimumreflection}) in the regime $\mathcal{E}(\eta)\gg1$. Equating this to the approximation (\ref{eq:simplifiedreflection}) we arrive at an equation which is very similar to the one obtained in the non-relativistic case. We find the appropriate laser frequency chirp, ensuring condition (\ref{eq:optimum}) to be satisfied in the regime $\gamma\gg1$ to be given by
\begin{align}\label{eq:URoptimum}
 \omega_\text{UR}^\text{opt}(\eta) = \frac{eE(\eta)}{m_e} \left(\frac{\left|eE^2(\eta)\right|-2m_i E'(\eta)}{\sqrt{2m_i E'(\eta) \left|eE^2(\eta)\right|}}\right).
\end{align}
By virtue of the condition $\mathcal{E}(\eta)\leq1$ established subsequent to eq.~(\ref{eq:optimummomentum}) we can rewrite condition (\ref{eq:derivativecondition}) to read $2m_i E'(\eta) \leq \left|e E^2(\eta)\right|$ and conclude that latter condition (\ref{eq:derivativecondition}) to be required for $\omega_\text{UR}^\text{opt}(\eta)$ to have positive, i.e., physical solutions. We can hence interpret that condition as the physical prerequisite of the optimally matched reflection coefficient to be reachable through tuning the frequency.
\begin{figure}[t]
  \begin{center}
    \includegraphics[width=\linewidth]{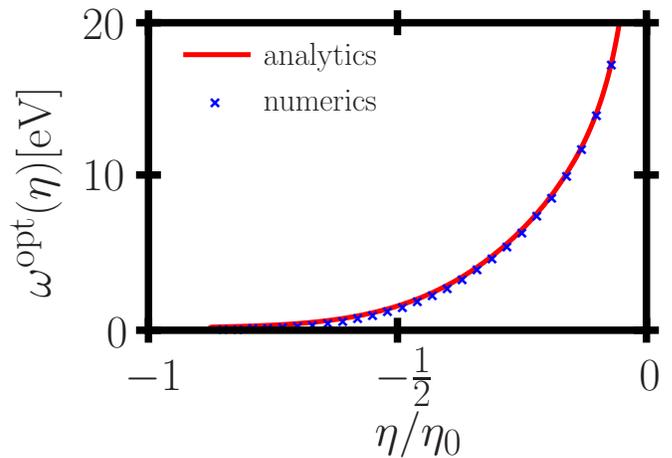}
    \caption{Optimal laser frequency in the ultra-relativistic model case (parameters in the text) according to the analytical solution eq.~(\ref{eq:URoptimum}) (solid red) in comparison to the numerical solution of eqs.~(\ref{eq:simplifiedreflection},\ref{eq:optimumreflection}) (blue crosses).}
    \label{fig:URoptimum}
  \end{center}
\end{figure}
Comparing the analytical expression (\ref{eq:URoptimum}) to the numerically consistent solution of eqs.~(\ref{eq:simplifiedreflection},\ref{eq:optimumreflection}) for a foil of $1$ nm thickness and density $n_e = 10^{23}\, \text{cm}^3 \approx 60\, n_\text{crit} $ accelerated by a laser pulse of intensity $I=10^{19}\, \text{W}/\text{cm}^2$ and duration $\tau_L = 3\times 10^4/\omega_0$ we find our analytical approximation very well confirmed (s. fig.~\ref{fig:URoptimum}). We also find the required laser frequencies to lie in significantly lower energy ranges as compared to the nonrelativistic case. This behavior can be explained by the observation that the equality between eqs.~(\ref{eq:simplifiedreflection},\ref{eq:UR_optimumreflection}) in the regime $\mathcal{E}(\eta)\gg1$ is achieved for larger values of $a_0$ than the equality between eqs.~(\ref{eq:simplifiedreflection},\ref{eq:NR_optimumreflection}) in the regime $\mathcal{E}(\eta)\sim 1$. Physically, this translates to the observation that an ultra-relativistic foil can withstand stronger laser acceleration, as experienced in lower frequency fields. Also the divergence of the optimum frequencies for later phases $\eta \to 0$ is readily explainable as the field derivative $E'(\eta)$ goes to zero in this regime, indicating that the optimum reflection coefficient (\ref{eq:UR_optimumreflection}) vanishes as well, which is achieved for very large frequencies, only. We furthermore note that at the time instant $\eta=0$ in the above example the foil has already been accelerated to $\gamma(\eta=0) = \mathcal{E}(\eta=0) \sim 10$, while constantly maintaining the optimum condition (\ref{eq:optimum}). This example indicates that indeed the suggested method is also applicable to ultra-relativistic foil motion, provided one can supply the required high-frequency photon beams.
\subsection{Conclusion}

In summary, we have presented a systematic study of how an appropriately chosen frequency chirp serves to maintain an optimum condition in the radiation pressure acceleration of a thin foil. We presented analytical expressions for the required pulse chirp in two limiting cases of nonrelativistic as well as ultra-relativistic foil motion. Comparing these limiting cases to exact numerical solutions of the defining equations of the pulse's frequency structure required to maintain optimal acceleration conditions we found excellent agreement between the exact and approximate solutions. While we found the required frequencies to be beyond the capabilities of nowadays available technology, the presented conceptual analysis may still prove useful for an improved understanding of the overall acceleration process.

SSB acknowledges support by Laboratory Directed Research and Development (LDRD) funding from Lawrence Berkeley National Laboratory provided by the Director, and the U.S. Department of Energy Office of Science, Offices of High Energy Physics and Fusion Energy Sciences under contract No. DE-AC 02-05 CH11231.

\end{document}